# Population-mobility coevolution drives the emergence of spatial heterogeneity in cities

Hao Huang[1], Yuming Lin[1*], and Jiazhen Liu[2*]

[1]Department of Urban Planning, Tsinghua University, Beijing, P.R.China.
[2]Department of Electronic Engineering, Tsinghua University, Beijing, P.R.China.
[*]Corresponding Authors

## ABSTRACT

The spatial heterogeneity of cities — the uneven distribution of population and activities — is fundamental to urban dynamics and related to critical issues such as infrastructure overload, housing affordability, and social inequality. Despite sharing similar scaling laws of population and mobility, cities exhibit vastly different spatial patterns. This paradox call for a mechanistic explanation for the emergence of spatial heterogeneity, while existing qualitative or descriptive studies fail to capture the underlying mechanisms. Here, we propose a coupled dynamical model that describe the intra-city population-mobility coevolution, explaining spatial heterogeneity as an emergent outcome of mutual feedback between the fast-changing mobility and the slow-adapting population. Our model is validated on over 388 million records from eight diverse global cities, successfully reproduces both the statistical laws and realistic spatial patterns. We find out realistic heterogeneity emerges as a distinct stable state, intermediate between disordered homogeneity and unsustainable super-hub dominance. Moreover, we theoretically and empirically show that populated areas are predominantly shaped by coevolution strength, while the increasing distance decay leads cities through a three-phase transition of homogeneity-heterogeneity-homogeneity. Besides, functional attractiveness between areas consistently enhances the ordered heterogeneous structure. Simulations of real-world planning scenarios – including crisis-induced lockdown, planned zone expansions, and dispersal from congested centers — indicate that integrated population-mobility policies are more cost-effective than single interventions. Our model can provides mechanistic, high-resolution insights to rigorously inform policy design.

Cities universally exhibit profound spatial heterogeneity in their internal functions, morphology, and social attributes, which reflects a bottom-up complexity and contrasts sharply with the orderly ideals of early urban planning[1,2]. In particular, the population distribution within cities is strikingly uneven: dense urban cores coexist with sprawling suburban zones, migrant enclaves emerge alongside long-established neighborhoods, and transient clusters of mobile populations form around transport hubs[3,4]. This heterogeneity does not stem from deliberate planning, but rather the dynamic interplay of human mobility and spatial concentration[5–7]. Understanding the mechanisms of heterogeneity is crucial for urban science, as it provides a systematic explanation of how population self-organize in urban complex systems. Theoretically, it can advance a unified framework to interpret the emergent spatial pattern of cities; practically, it can equip planners and policymakers with predictive tools for addressing pressing challenges such as congestion, segregation, and sustainable resource allocation[8,9].

Urban scientists have made significant progress in characterizing the urban complex systems, with major efforts devoted to urban growth patterns[10,11], mobility pattern[12] and demographic dynamics[10,13,14] through natural increase, migration, and rank–size dynamics of urban systems. Although these works have achieved notable success in explaining growth processes and inter-city migration, they usually treat cities as homogeneous wholes and therefore fall short in addressing the intra-city spatial heterogeneity. More recently, a few studies have examined the spatial structure of intra-city population growth[15,16], yet they either model population as a statistical aggregation or provide only descriptive correlations, without clarifying the mechanisms through which ordered internal structures emerge.

Another large body of research has focused on modeling intra-city mobility. Classical work such as the gravity model and its many variants[17–19] assume that the mobility flows is the function of population, capturing aggregate mobility patterns across the system. The recently radiation model and other stochastic flow models[6,20] introduce parameter-free opportunity mechanisms and successfully reproduce large-scale migration and commuting patterns. However, a pervasive assumption in these literature is that, the underlying population distribution is a static background when focusing solely on the allocation of mobility. This simplification overlooks a critical fact: although population redistribution evolves more slowly than mobility dynamics, it is fundamentally coupled with mobility. Here, a core insight from complex systems theory becomes relevant[21]: in systems with coupled fast and slow dynamics, the slow variable, though evolving gradually, actively constrains the dynamics of the fast subsystem[22,23]. Freezing the slow variable artificially removes the constraint and, therefore, leads to fundamentally loss of the capability to account for the emergence of heterogeneous behavior[24–26]. As a result, models that only aggregate mobility cannot fully explain the persistent spatial heterogeneity observed within cities. Hence, a clear knowledge gap therefore persists in urban science: the absence of a unified dynamical framework capable of explaining the emergence of dynamically stable spatial heterogeneity within cities.

The purpose of this paper is to uncover the dynamic mechanism driving the emergence of spatial heterogeneity in cities. We develop a nonlinear dynamical model that allows the slow population dynamics to coevolve with the fast human mobility dynamics. We analytically and numerically find that our coevolution model successfully accounts for the spatial heterogeneity of the real world, validated by datasets from eight cities worldwide with about 388 million data records. Our theoretical approach shows that the absence of coevolution strength leads to trivial homogeneity inside the city. As the increasing of coevolution strength between the population and mobility dynamics with appropriate distance decay and functional attractiveness, the heterogeneity within cities emerges and eventually reach the heterogeneity we observe in the real cities. Moreover, we also find that the extreme coevolution strength will finally drive the city phase further to abnormal super-hub-dominated heterogeneity, which is theoretically predicted in our methods. Three real urban planning scenarios verify the practical value of our population-mobility coevolution framework. In summary, our dynamical system offers a generative, mechanistically transparent explanation, providing a quantifiable scientific foundation for scenario modeling and policy intervention. Unlike classical flow-allocation models, our framework captures how endogenous coevolution between mobility and population shapes the emergence of spatial heterogeneity.

## Empirical Observations

To examine the spatial heterogeneity of the population in real cities, we used large-scale datasets on human mobility and population. Our analysis is based on a comprehensive dataset comprising 261 millions of mobility records, supplemented by 87,932,141 population counts and 3,938,754 points-of-interest (POI) data from eight cities in China, the United States, Brazil, and Japan. These datasets, originally at varying spatial and temporal resolutions[27,28], were harmonized and aggregated to a uniform spatial scale of 2 km. Detailed descriptive statistics of these datasets are provided in supplementary material.

Interestingly, the probability distributions of population and mobility flows in these cities exhibit remarkable similarity (Fig. 1a), a phenomenon widely recognized as the urban scaling law. However, there are obvious differences in geographic setting, land area, polycentricity and population size in the cities, which manifest as strong heterogeneity in the spatial distributions of population (Fig. 1b). For instance, Beijing, on the North China Plain, is predominantly monocentric, with more than 20 million residents concentrated around a single core across about 16,411 $km^2$; Rio de Janeiro, constrained by coastal mountains and the Atlantic Ocean, accommodates nearly 6 million within an almost continuous coastal urban fabric of about 1,200 $km^2$; Santa Clara County, in California's coastal valley, covers approximately 3,344 $km^2$ and is distinctly polycentric, with about 2 million residents distributed across multiple suburban nodes. Similar statistical regularities amid pronounced spatial heterogeneity, such contrast presents a striking paradox and naturally raises a scientific question: Is there a universal mechanisms underlying city spatial heterogeneity?

To address this paradox, there is a great need for an endogenous mechanism through which mobility and population can converge toward similar statistical regularities while preserving distinct spatial heterogeneity. In essence, the two evolve on different timescales; yet intuitively, fast-changing mobility patterns gradually reshape the slow-adapting population landscape, while the emergent population distribution, in turn, constrains subsequent mobility behaviors. Building on this intuition, we develop a population–mobility coevolution framework that explicitly captures these coupled dynamics.

## Slow-fast population-mobility coevolution framework

Here we explicitly model the coevolution between regional population distribution and inter-regional human mobility. In contrast to traditional flow assignment models that treat population distribution as exogenous, our model treats both mobility and population as endogenous state variables, evolving jointly over time. The system consists of two interdependent components: (i) Mobility flow dynamics, describing how people move between regions under the influence of population in regions; (ii) Population dynamics, capturing how these flows reshuffle the population distribution. Crucially, these two dynamics are mutually coupled in a feedback loop: mobility flows respond to the current population landscape, and in turn, the net flows between areas reshape future population states. This mutual dependency defines a coevolutionary mechanism in cities.

Our model is therefore structured as a fast–slow dynamical system, where population and mobility coevolve over different timescales. Specifically, mobility flows adjust rapidly in response to changes in population distribution (fast variables), while population evolve more gradually through cumulative net flows (slow variables). This separation of timescales follows the classical logic of the Born-Oppenheimer approximation and has been widely applied in fast-slow coevolution models[21,23].

Let $x_i(t)$ denote the population of region $i$, and $\Pi_{ij}(t)$ denote the instantaneous flow propensity from region $j$ to $i$. We first define the mobility rate as:

$$w_{ij}(t) \sim r_{ij}^{\alpha} \theta_{ij}^{\beta} x_i^{\lambda}(t), \tag{1}$$

where $r_{ij}$ is the inter-regional distance, $\theta_{ij}$ denotes the functional similarity between regions $i$ and $j$ (e.g., based on POI distributions), and $\lambda$ controls the coevolution strength.

At each time step, this generates a dynamic OD matrix $W(x,t)$, which in turn reshapes the population states according to:

$$\frac{dx_i}{dt} = \sum_j [w_{ji}(x,t) - w_{ij}(x,t)]. \tag{2}$$

This equation ensures quasi-conservation of population via net mobility flux.

To characterize the aggregate dynamics, we derive a continuous limit of the system, where the probability density $P(x,t)$ of regional populations evolves as:

$$\frac{\partial P(x,t)}{\partial t} = -\frac{\partial}{\partial x}\left[\int (W(x',t) - W(x,t))P(x,x',t)dx'\right], \tag{3}$$

Simultaneously, the joint distribution $P(x,x',t)$ evolves through a birth-death process, where its stationary state will satisfy ($t \to \infty$):

$$P(x,x') = W(x)P(x)P(x'), \tag{4}$$

This captures the evolution of flow structure driven by population states.

Finally, the flow weight $w_{ij}(t)$ between regions evolves toward a target value $\mu(x,r,\theta) = x^{\lambda} r^{\alpha} \theta^{\beta}$, with its distribution following:

$$\frac{\partial P(w,t)}{\partial t} = -\frac{\partial}{\partial w}[(\mu(x,r,\theta) - w)P(w,t)], \tag{5}$$

a drift-type Fokker–Planck equation describing how actual mobility flow converges to its dynamic attractor.

Together, Eqs. 1–5 define a fast-slow coevolutionary system that integrates mobility and population evolution with adaptation between them. This formulation allows us to analytically and numerically investigate the emergence of spatial heterogeneity from first principles. Unlike models where flows are allocated based on fixed trip totals or exogenous population distribution, our framework dynamically generates flows endogenously from the current population configuration, and these flows actively reshape the population landscape, leading to long-term spatial heterogeneity. In this sense, our model belongs to the class of generative urban dynamical systems, rather than predictive assignment schemes. This feedback structure allows us to explore how microscopic mobility decisions aggregate into macroscopic spatial patterns, and to explain why cities with similar scaling laws can nonetheless exhibit vastly different internal heterogeneity.

## Emergence and transitions of spatial heterogeneity

Incorporating Eq. (3) with (5) allows us to solve the the steady-state probability distributions of population $P(x, t \to \infty)$ and the intensity of mobility flows $P(w, t \to \infty)$ under fast-slow coevolutionary dynamics, analytically finding that they satisfy (see Method M2 for analytical details) $P(x) \sim x^{-\lambda'}$ and $P(w) \sim w^{-\gamma}$. Specifically, the scaling exponent of population distribution within a region follows $\lambda < \lambda' < 2\lambda$, where $\lambda$ is the coevolution strength exponent. This analytical result is validated by the real observations (see SM Table S2). To further validate our coevolution model, we compared our numerical results to the empirical data for eight different cities. Figure. 2 shows that the best-fitted modeling results of $P(x)$ and $P(w)$ of eight cities are consistent with real-city observations. We report our simulation methods and best-fitted parameters in Method and Supplementary Materials.

More strikingly, our model also captures the patterns of spatial heterogeneity (Fig. 2) for eight cities that exhibit totally different spatial distributions of population. Moreover, the conditional expectations $E[W|X_o,X_d]$ also closely match empirical observations (Fig. 3). This indicates that the coevolution model can mechanistically generate city spatial heterogeneity as well as the probability distribution of population and mobility.

To further answer the necessity of these mechanism in the occurrence of heterogeneity, we explore spatial heterogeneity under widely different parameters. The simulation shows that the population heterogeneity of city undergoes distinct phase transitions (Fig. 4), modulated by the parameters of coevolution strength $\lambda$, distance decay $\alpha$, and functional similarity $\beta$.

We identify four characteristic states: Homogeneity (state 1), Quasi-homogeneity (state 1.5), Realistic heterogeneity (state 2), Superhub-dominance heterogeneity (state 3).

With increasing coevolution exponent $\lambda$ ($\alpha = 1.4, \beta = 3$), three distinct patterns emerge (Fig. 4a). A state of homogeneity (state 1) is captured in the city when $\lambda = 0$, with impulse-like population probability distributions. That implies without population-mobility coevolution, the population in the city is like the sand brushed over a tabletop. At $\lambda = 0.5$, realistic urban heterogeneity appears (state 2), with both population and flows following power-law distribution. This confirms that population-mobility coevolution is indispensable for the formation of spatial heterogeneity. When coevolution strength increases further, multi-centered structures collapse toward a single superhub (state 3), diverging from the real city case. Such concentration exceeds real land capacity, indicating that only moderate coevolution strength maintains stable ordered structures. Across all cities that we examine, the estimated coevolution strength lies in the interval $[0.44, 0.50]$. Higher coevolution strength indicates stronger self-organization and agglomeration effects in densely populated areas (e.g., Beijing, $\lambda = 0.5$; Guangzhou, $\lambda = 0.5$), while less aggregated or smaller cities are associated with lower adaptive coevolution strength (e.g., Santa Clara County, $\lambda = 0.44$; Rio de Janeiro, $\lambda = 0.44$).

A non-monotonic process where heterogeneity first strengthens and then diminishes (Fig. 4b) is observed, with increasing distance-decay exponent $\alpha$ ($\beta = 3, \lambda = 0.5$). Taking Guangzhou(China) as an example, when $\alpha = 0$, the spatial distribution of population approaches quasi-homogeneity (state 1.5), with exponential trip-length distributions and Gaussian-like population distributions. At $\alpha = 1.4$, simulations achieve optimal fit with empirical cities, exhibiting realistic spatial heterogeneity (state 2), where both population and mobility follow power-law distributions. The population gradually gather and live in the center of Guangzhou, forming an ordered structure with dynamic stability. When $\alpha$ increases further, the system reverts to strong homogeneity (state 1). The population distributes uniformly, with impulse-like probability distributions, reflecting high travel costs in the city because of low transport accessibility or strict community control. We further observe that most cities have distance-decay exponents $\alpha \in [1.4, 1.7]$, below the empirical benchmark of 2, reflecting the convenience of public transport and other travel modes beyond 2 km. A higher decay exponent (e.g., Shanghai, $\alpha = 1.7$; San Antonio, $\alpha = 1.7$) indicates lower willingness for long-distance trips, likely due to high-quality local services that satisfy most needs.

With increasing functional-similarity exponent $\beta$ ($\alpha = 1.4, \lambda = 0.5$), we observe two distinct patterns (Fig. 4c). At $\beta = 0$, cities display homogeneity (state 1), with mobility following a Poisson distribution. At $\beta \geq 3.0$, the population distribution converges to a stable ordered structure in Guangzhou. The functional-similarity exponent $\beta$ captures preferences in travel destinations, with lower values (e.g., Shanghai, $\beta = 1.6$; Santa Clara County, $\beta = 1.7$) indicating a stronger tendency to explore and revisit heterogeneous urban areas, whereas higher $\beta$ suggests clustering around functionally similar locations (e.g., Beijing, $\beta = 4$; Rio de Janeiro, $\beta = 3.2$). To further assess the role of functional heterogeneity, we compute Shannon entropy of POI distributions across four cities. Fig. 4d shows that high-entropy regions coincide with population clusters, suggesting that functionally diverse areas attract more population settlement.

## Planning and policy implications

In practical planning and policymaking, situational analysis provides only the starting point, and the critical challenge lies in realizing the envisioned goals through facility investment, population attraction, and other diverse policies. The dynamical model developed in this study supports scenario-based simulations, providing forecast under alternative planning or policy scenarios. We illustrate its applicability through three representative policy scenarios: (i) short-term lockdown under epidemic, (ii) population growth in newly planned development zones, and (iii) population dispersal from congested city centers. These scenarios align with archetypal urban planning and policy challenges commonly encountered by large metropolitan areas.

The first scenario simulates the Shanghai lockdown from March to May 2022, implemented to contain the spread of COVID-19 and facilitate nucleic acid testing. At that time, the government divided the city along the Huangpu River, enforcing lockdown in separated phases with no cross-river mobility, accompanied by strict community-level restrictions. We represent these policy constraints by (a) prohibiting cross-bank OD connections and (b) increasing the distance-decay exponent $\alpha$ to reflect higher travel costs under restrictions. As shown in Fig. 5a, the total OD flows declined by 83.1% relative to the pre-lockdown baseline. The joint distributions of population and mobility reveal that flows between high-density zones dropped sharply, while the overall spatial distribution of population remained largely unchanged. These findings are consistent with reality, underscore the relative vulnerability of human mobility under crisis restrictions, suggesting more caution exercised before action.

The second scenario examines population growth in Guangzhou's Nansha New Area. According to the *Guangzhou Nansha Master Plan (2021–2035)*, Nansha aims to be a subcenter of Guangzhou and as a hub for marine science and innovation, targeted to reach a population of 2 million by 2035. Based on the vision proposed in the master plan, we designed three sub-scenarios: (a) infrastructure investment only — adds 16 facilities per 1 km$^2$ within the designated development boundary, covering research institutes, industrial enterprises, residential services, and amenities; (b) talent policy only — increase the coevolution parameter $\lambda$ by 0.021, reflecting residents' retention willingness under talent policy; and (c) a combined approach

— adding 16 facilities per km$^2$ while increasing $\lambda$ by 0.012. Fig. 5b (columns 3–5) shows that in sub-scenario (a), despite the addition of numerous multi-functional facilities, the effect on population growth was limited because of the district's peripheral location relative to the city core. In sub-scenario (b), talent policies proved attractive in facility-rich areas but were ineffective in poorly serviced zones, failing to generate the three major clusters envisioned in the master plan. In sub-scenario (c), the combined policy achieved the target population of 2 million and created continuous clusters of settlement, demonstrating that infrastructure and talent policies must be implemented in synergy to realize effective population growth in new development zones. These findings highlights the value of infrastructure-population integrated approaches in achieving targeted expansion in peripheral zones.

The third scenario evaluates population dispersal from Beijing's central six districts. The *Beijing Master Plan (2016–2035)* sets a target of reducing central population to below 10.85 million, by redistributing non-capital functions to the sub-center and peripheral areas. Based on measures proposed in the master plan, we simulate three sub-scenarios: (a) infrastructure expansion only, with 1 facility for each categories (government, finance, research, residential, and services) in non-core km$^2$ grids; (b) talent policy only, where residents' retention willingness in non-core areas is increased by $\lambda + 0.01$; and (c) a combined strategy, with each km$^2$ adding 0.5 facilities and $\lambda + 0.005$. Fig. 5c (columns 3–5) shows that in sub-scenario (a), suburban infrastructure development relocated about 0.7 million residents out of the six central districts. In sub-scenario (b), talent policy alone proved more effective in dispersing population, compared to relying solely on infrastructure. In sub-scenario (c), the combined measures relocated about 1.6 million residents with lower levels of infrastructure investment and talent subsidies. These findings imply that coordinated measures can achieve comparable dispersal outcomes at substantially lower cost, highlight the importance of population-mobility coevolution, especially for cost-effective population dispersal in densely populated metropolises.

Collectively, these real-world scenario highlight that the proposed dynamical model can be used to evaluate planning interventions outcomes quantitatively. In our simulations, parameters for facility investment are derived from population targets in master plans, while adjustments to talent-attraction policies are represented as proportional increases to the existing coevolution strength. By parameterizing planning initiative and policy in the model, it provides simplified but effective analysis method for designing targeted, cost-effective policies in both short-term shocks and long-term structural adjustments of cities.

## Discussion

This study presents a dynamical framework that explicitly models the coevolution between regional population and mobility flows, revealing a fundamental mechanism driving the emergence of spatial heterogeneity. While prior models, such as gravity, radiation, and other flow allocation schemes, have advanced our understanding of intra-city mobility patterns, they generally assume population distribution as a static background. Our work departs from this tradition by treating both population and mobility as endogenous, time-evolving variables, governed by a mutually coupled fast-slow dynamical system.

This distinction is not merely semantic. Drawing from insights in complex systems theory and non-equilibrium statistical physics, we show that even when population evolves slowly, its coupling to mobility flows plays a decisive role in shaping long-run urban structure. Freezing the slow variable (population) removes critical dynamical constraints and fails to reproduce the observed persistence and diversity of spatial heterogeneity across cities. In this sense, our model goes beyond traditional models: it captures how small-scale mobility decisions, interacting with structural distance and functional affinity, self-organize into stable macroscopic population patterns.

A key discovery from our model is that realistic urban heterogeneity represents a distinct phase, analogous to a critical state in complex systems[29]. As the coevolution strength $\lambda$ increases, the urban system transitions from a disordered, gas-like 'quasi-homogeneity' to an ordered, crystal-like 'super-hub-dominance'. The complex, polycentric structure of real cities exists in a delicate balance between these extremes, a state of 'realistic heterogeneity' that is dynamically stable yet highly adaptive. This finding resonates with theories of cities as complex systems poised at a state of self-organized criticality, balanced between order and chaos to foster resilience[30,31]. This critical state implies that urban systems are sensitive to perturbations; policies can either stabilize this productive heterogeneity or inadvertently push the city toward inefficient homogeneity or unsustainable concentration[32]. Our model provides a mechanistic explanation for this delicate balance and offers a tool to navigate the trade-offs inherent in maintaining urban vitality.

Our framework represents a significant methodological advance for spatial heterogeneity, shifting the focus from correlational description to causal, mechanistic explanation. Unlike descriptive techniques such as spatial clustering or statistical analysis[33–35], which identify *what* patterns exist, our generative model explains *why* they emerge from first principles. Critically, our model's success extends beyond fitting statistical distributions; it reproduces the actual geographic patterns of population distribution, addressing the paradox that cities with similar statistical distributions can have vastly different spatial heterogeneity.

Consequently, our model serves as a powerful tool for policy evaluation, avoiding the causal biases inherent in static model predictions. It provides practical solutions and predictive capabilities for key challenges in urban planning and policy, including the establishment of new urban zones, deconcentration of population in high-density districts, and interim control measures during crises—challenges. In this study, we simulated multiple policy instruments across three scenarios, encompassing infrastructure provisioning, talent policies, and their integrated applications. Notably, the combined effects of infrastructure provisioning and talent policies are more effective and cost-efficient than single interventions. Furthermore, our model delivers mechanistic explanations and high-spatial-resolution population changes after policy enactment, which offers evidence-based guidance for urban planning and policymaking.

While this study provides a foundational framework, it is not without limitations, which in turn open promising avenues for future research. The model's fit was less precise in some low-density US areas and high-density Japanese areas, potentially due to data noise, suggesting opportunities for refinement with richer datasets. Looking forward, a natural extension is to incorporate demographic dynamics of population growth, such as births, deaths, and migration[10,11,36], into the coevolutionary framework. This would allow for the simulation from initial settlement to mature metropolitan structures. Furthermore, future work should explore the coupling of population and mobility with other critical urban elements[37,38], such as land use, housing markets, and transportation infrastructure[39], to build more holistic, multi-layered models of urban coevolution. Such efforts will be crucial for developing a comprehensive science of cities that can effectively guide sustainable and equitable urban development.

## Methods

### M1:Data

To examine population spatial heterogeneity in real cities, we employ large-scale datasets of human mobility and population. The first dataset is obtained from one of the largest mobile service providers in China, covering four Chinese megacities with one-week trajectories in 2023–2024 at a temporal resolution of 30 minutes and a spatial resolution of 200–250 m. The second dataset comes from one of the largest mobile service providers in U.S. in 2019, providing one month of mobility traces for two U.S. cities at 30-minute intervals and census-tract resolution. Human mobility data for Rio de Janeiro are drawn from an open dataset[27], covering the entire year of 2014 with daily resolution at the census-tract level. A further dataset for a major anonymous Japanese metropolitan area includes mobility, population, and POI data with 30-minute resolution at 500 m grids[28]. In addition, the WorldPop dataset provides gridded population counts for non-Chinese cities, while POI data are sourced from Amap for Chinese cities and from Overture Maps for U.S. and Brazilian cities. All datasets are harmonized and aggregated to a uniform spatial scale of 2 km.

### M2:Analytical framework of population-mobility coevolution

*Probability distribution of population at regions*

The goal of this section is to derive the probability distribution of population $P(x)$ shown in the main text. We begin with a simple quasi-continuity balance of human mobility [15], that is, the population change in regions approximately equal to the mobility flux of cities. This actually assumes the population of a city does not change intensively in a relatively short time period, similar to the quasi-equilibrium state in statistical physics. Incorporating the (1) with (2) leads to the continuity equation of Eq. (3), whose steady-state satisfies

$$\int_0^{x_{max}} (x'^\lambda - x^\lambda) P(x,x') dx' = C/\kappa, \tag{6}$$

where $C$ is a non-negative constant, and $\kappa = \int_0^{r_{max}} dr \int_{-1}^1 d\theta r^\alpha \theta^\beta P(r) P(\theta)$. Specifically, $C = 0$ suggests that there is no mobility flow in the urban system, which is obviously wrong. A more reasonable solution will be $C > 0$. In this case, urban systems show "steady flow equilibrium". Due to the mobility flow between regions, $x$ and $x'$ are not independent. Given (**??**), the evolution of $P(x,x',t)$ satisfies

$$\frac{\partial P(x,x',t)}{\partial t} = \kappa x^\lambda(t) P(x,t) P(x',t) - \gamma_0 P(x,x',t), \tag{7}$$

where $\gamma_0$ is a constant characterize the breaking probability rate of the mobility flow between two regions. Hence, solving the steady state of Eq. (7) leads to

$$P(x,x') = \kappa x^\lambda P(x) P(x') / \gamma_0. \tag{8}$$

Substituting Eq. (8) into (6) finds

$$P(x) = \frac{C_0}{\kappa |<x'^\lambda> - x^\lambda|x^\lambda}. \tag{9}$$

When $x$ is large, $x^\lambda \gg <x'^\lambda>$, hence $P(x) \sim x^{-2\lambda}$. On the other hand, when $x$ is small, $x^\lambda \ll <x'^\lambda>$, $P(x) \sim x^{-\lambda}$. Therefore, we can analytically predict that population distribution $P(x)$ asymptotically satisfies

$$P(x) \sim x^{-\eta}, \lambda < \eta < 2\lambda. \tag{10}$$

### *Mobility population distribution between regions*
Equation (1) leads to following dynamic equation of mobility population

$$\mathbf{W}'(t) = \mathbf{W}(t)\Pi. \tag{11}$$

For the mobility population $w_{ij}$ between regions $i$ and $j$, Equation (11) suggests

$$\frac{dw_{ij}}{dt} \propto (x_i^\lambda \theta_{ij}^\alpha r_{ij}^\beta - \gamma_0 w_{ij}). \tag{12}$$

Therefore, the probability of the average mobility population between regions at time $t$ satisfies

$$\frac{\partial P(w,t)}{\partial t} = -\frac{\partial}{\partial w}[P(w,t)(x^\lambda \theta^\alpha r^\beta - w)] \tag{13}$$

Solving Eq. (13) leads to following stationary solution

$$P(w|r,\theta,x) \propto \frac{1}{|\mu(x,r,\theta) - w|}. \tag{14}$$

The mobility population distribution between regions should follow

$$P(w) = C \int \int \int \frac{P(x)P(r)P(\theta)}{|\mu(x,r,\theta) - w|} dr d\theta dx. \tag{15}$$

### **M3:Simulation**
### *Simulation Procedures*
We propose a co-evolutionary dynamical model to simulate the interplay between population distribution and human mobility patterns. The simulation begins by initializing the system with a uniform population distribution across all spatial nodes. The model then proceeds in discrete time steps, iteratively updating the population of each node and the mobility flows between them.

At each step, the population of a given node evolves based on the net flux—the difference between total population inflow and outflow—calculated from the mobility network of the previous step. Subsequently, the mobility flow from node $i$ to node $j$ is recalculated as a function of three key components: the geographical distance ($d_{ij}$), the functional attractiveness ($s_{ij}$), and the destination node's population ($n_j$). This iterative process creates a feedback loop where population distribution shapes mobility, which in turn reshapes the population distribution. The simulation continues until the system reaches a steady state, identified by minimal changes in the population vector and a high degree of stability in the mobility matrix between consecutive steps.

### *Parameter Settings*
The model's behavior is governed by a set of key parameters. The distance decay exponent, $\alpha$, modulates the friction of distance. The similarity attraction exponent, $\beta$, determines the preference for connecting to functionally similar nodes. The preferential attachment exponent, $\lambda$, controls how population size attracts inflow. For the results presented, we systematically explored the parameter space by setting $\alpha \in [0, 3.0]$, $\beta \in [0, 4.0]$, and $\lambda \in [0, 1.0]$. The simulation proceeds with a time step of $\Delta t$ for up to $T = 8000$ iterations. The validity of each parameter set is assessed by comparing the resulting steady-state distributions of population and mobility flow against empirical data, using the Jensen-Shannon (JS) divergence as the primary goodness-of-fit metric.

## Code availability

The code and data used in this research is available at https://github.com/THU-EricHuang/Slow-fast-population-mobility-coevolution-model_v1.

## Acknowledgments

We thank Professor Lei Dong and Professor Chaoming Song for their valuable suggestions that improved the manuscript.

## Author contributions

All authors conceived the project; Y.L. and J.L. proposed the theoretical framework; H.H collected the data and performed the empirical analysis with help from Y.L.; J.L. conducted the theoretical analysis; H.H. and Y.L. performed the numerical analysis; all authors discussed and interpreted results; all authors wrote the manuscript.

## Competing interests

The authors declare no competing interests.

## Additional information

Supplementary information is available for this manuscript.

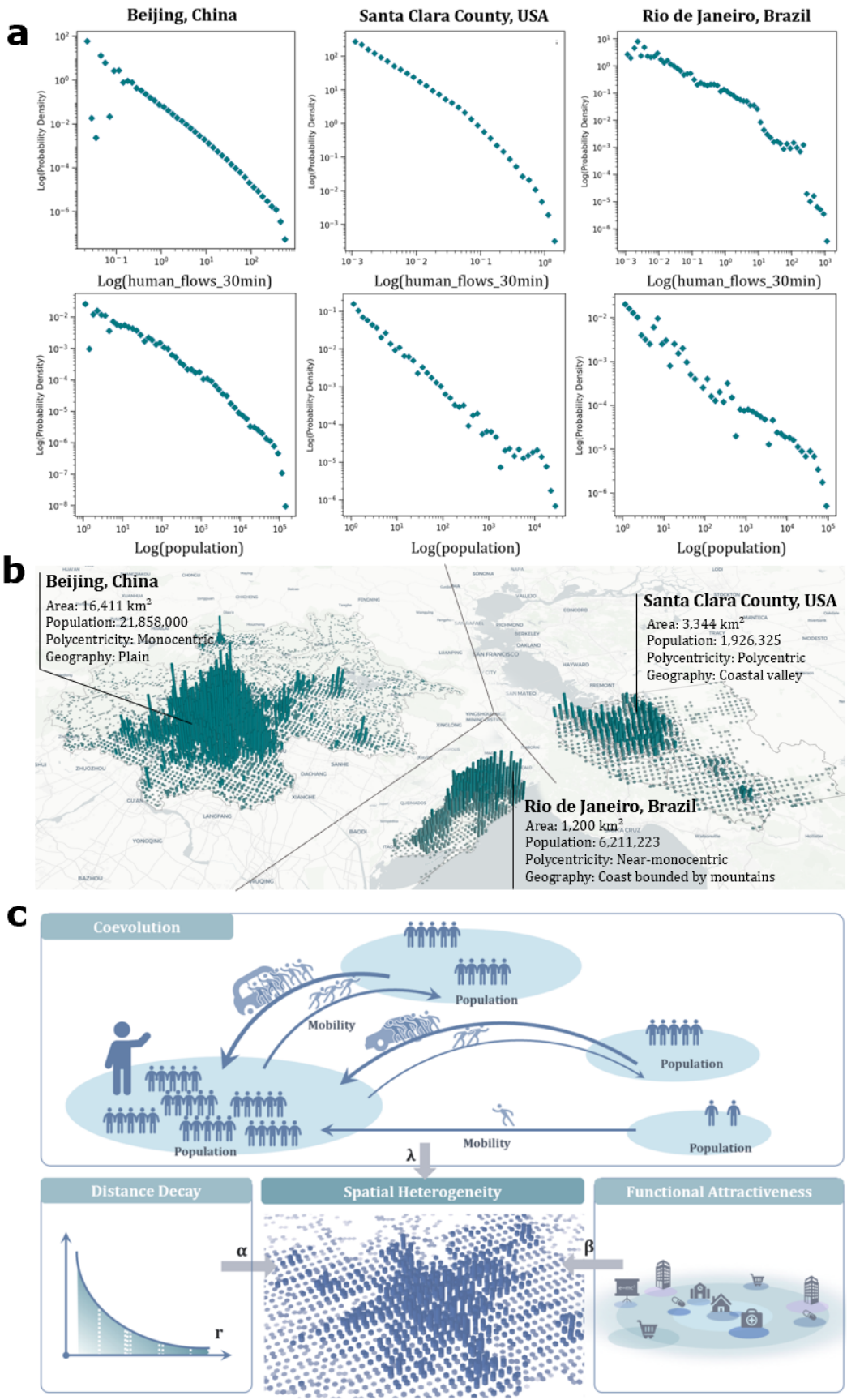

**Figure 1. Empirical observations and our proposed coevolution dynamics model. a**, The probability distributions of population and mobility flows in Beijing, Rio de Janeiro and Santa Clara County. **b**, The 3-D graph of the three cities with differences in geographic setting, land area, polycentricity and population size. **c**, Overview of an population-mobility coevolution model, where population and mobility influence each other to form spatial heterogeneity. Distance-decay and functional attractiveness mechanisms are included.

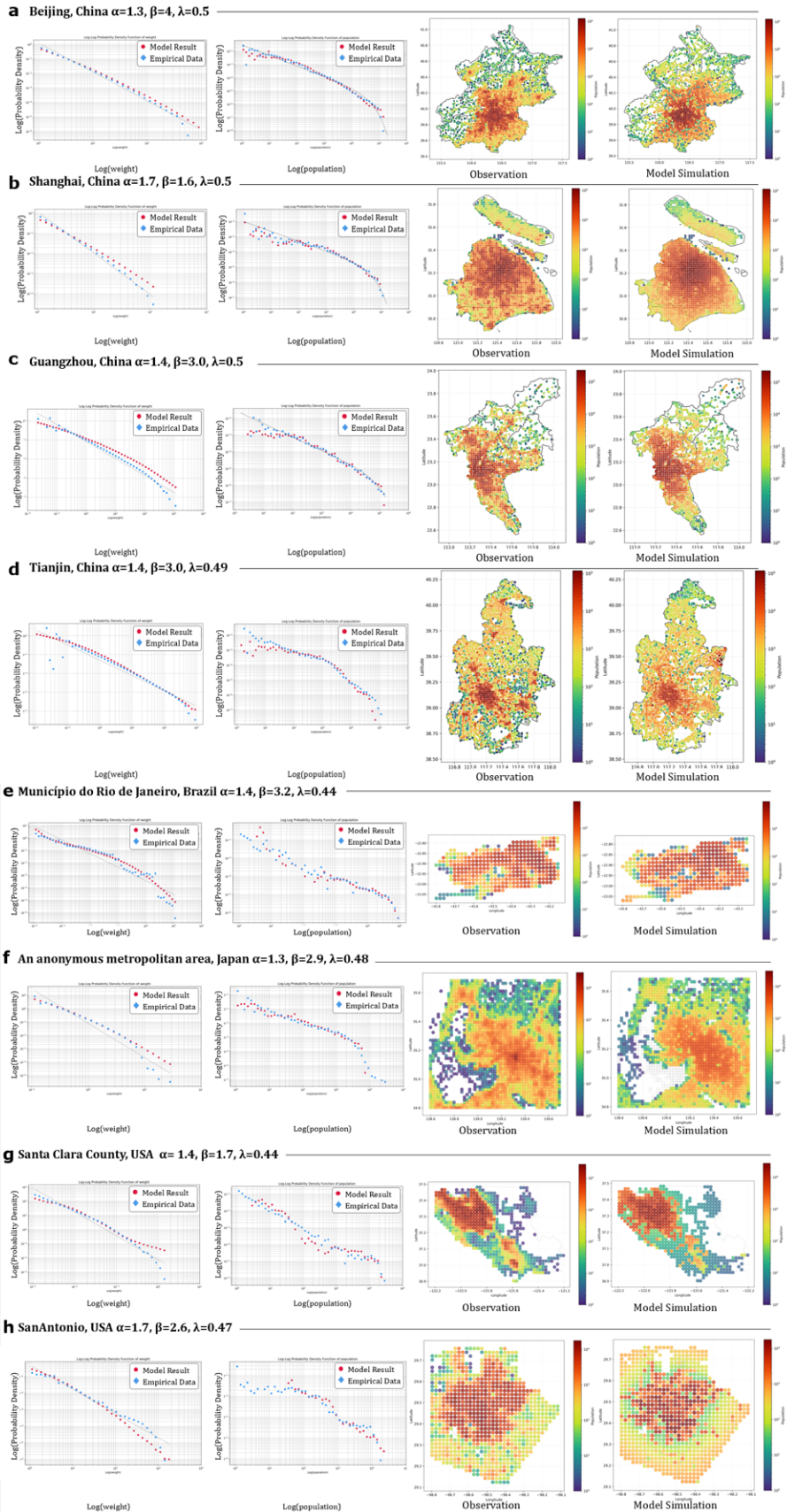

**Figure 2. Empirical data and the simulation results of the coevolution model. a-h**, Cases in Beijing (**a**), Shanghai(**b**), Guangzhou(**c**), Tianjin(**d**), Rio de Janeiro(**e**), an anonymous metropolitan area in Japen(**f**), Santa Clara County(**g**), SanAntonio(**h**). The first and second columns are the comparison of probability distribution of human mobility and population between empirical data and the simulation. The third and forth columns are the comparison of population spatial distribution between empirical data and the simulation.

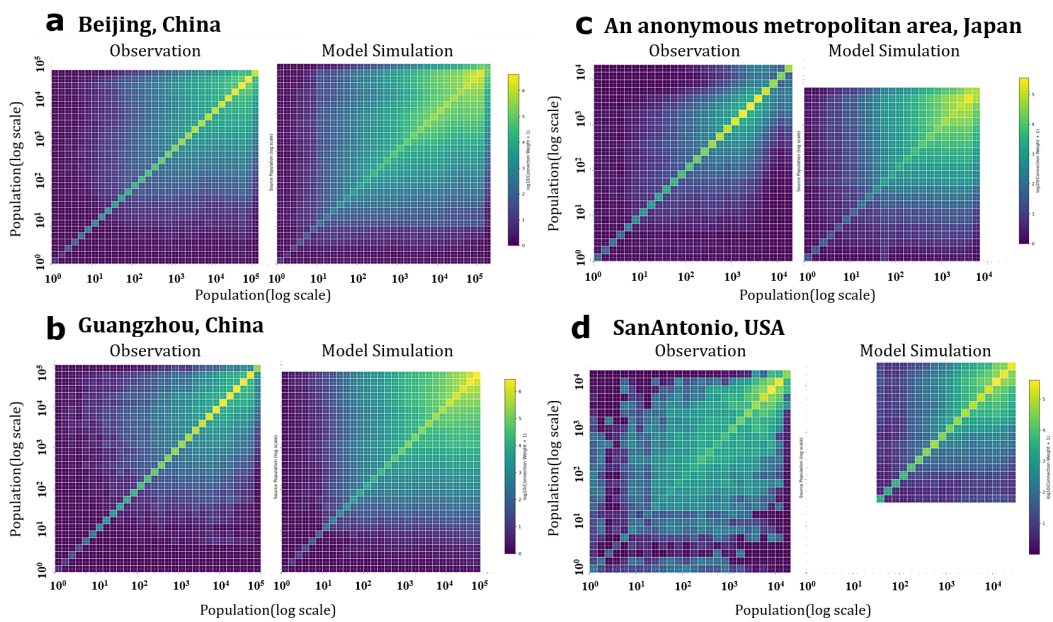

**Figure 3. Comparison of empirical and modeled mobility matrices. a-d**, Heatmap analysis for Beijing(**a**), Guangzhou(**b**), an anonymous metropolitan area(**c**) and San Antonio(**d**). Heatmaps illustrating the mobility connections between populations, with nodes ordered by population size along the axes. The color intensity represents the strength of mobility flows between nodes.

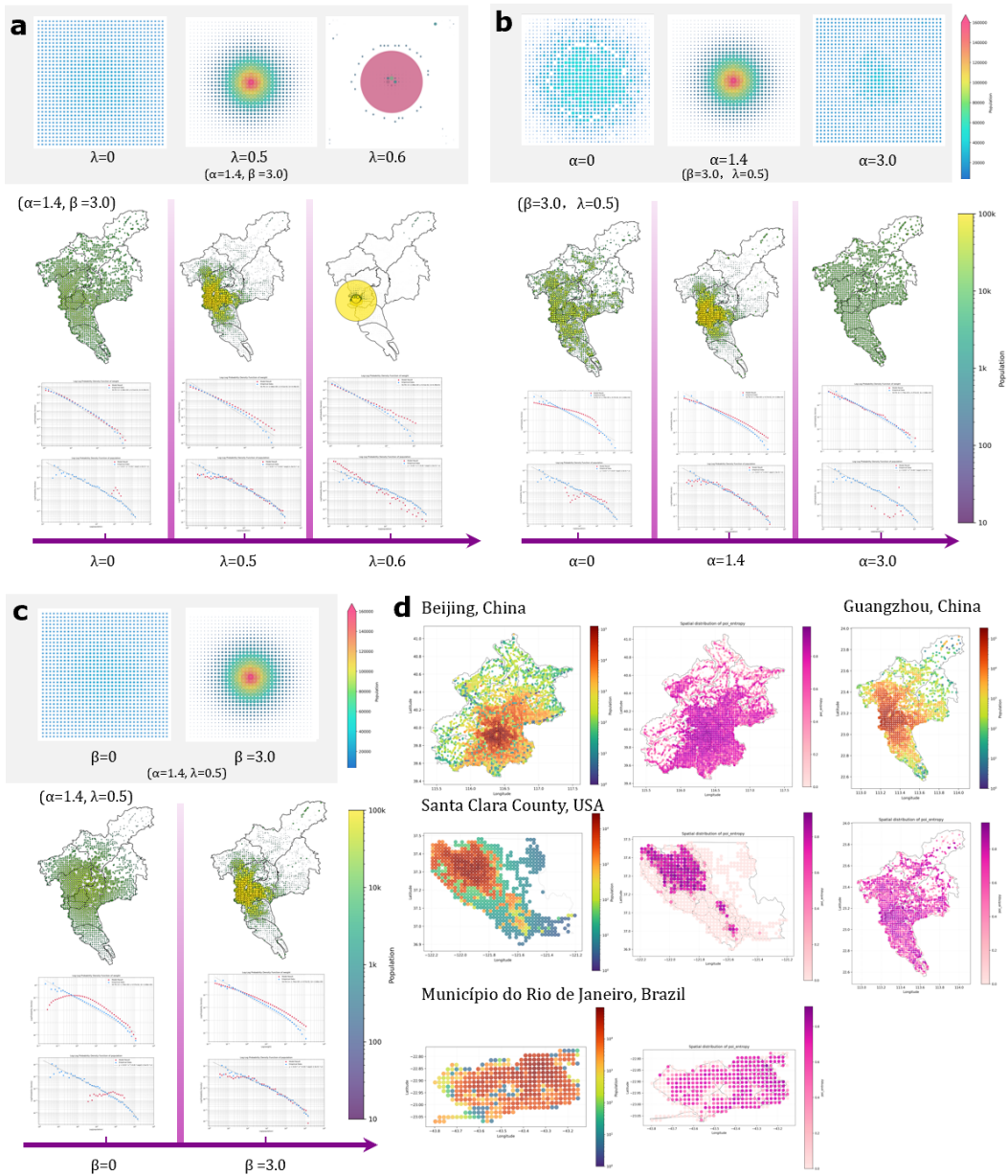

**Figure 4. Phase diagram and Shannon entropy of multiple-category POIs. a-c**, Phase transition, statistical distribution of population and human mobility, and spatial distribution of population under different $\lambda$(**a**), $\alpha$(**b**) and $\beta$(**c**). The upper is the phase transition of toy model. The lower is the phase transition simulation of case in Guangzhou(China). **d**, The comparison of population spatial distribution and Shannon entropy of multiple-category POIs.

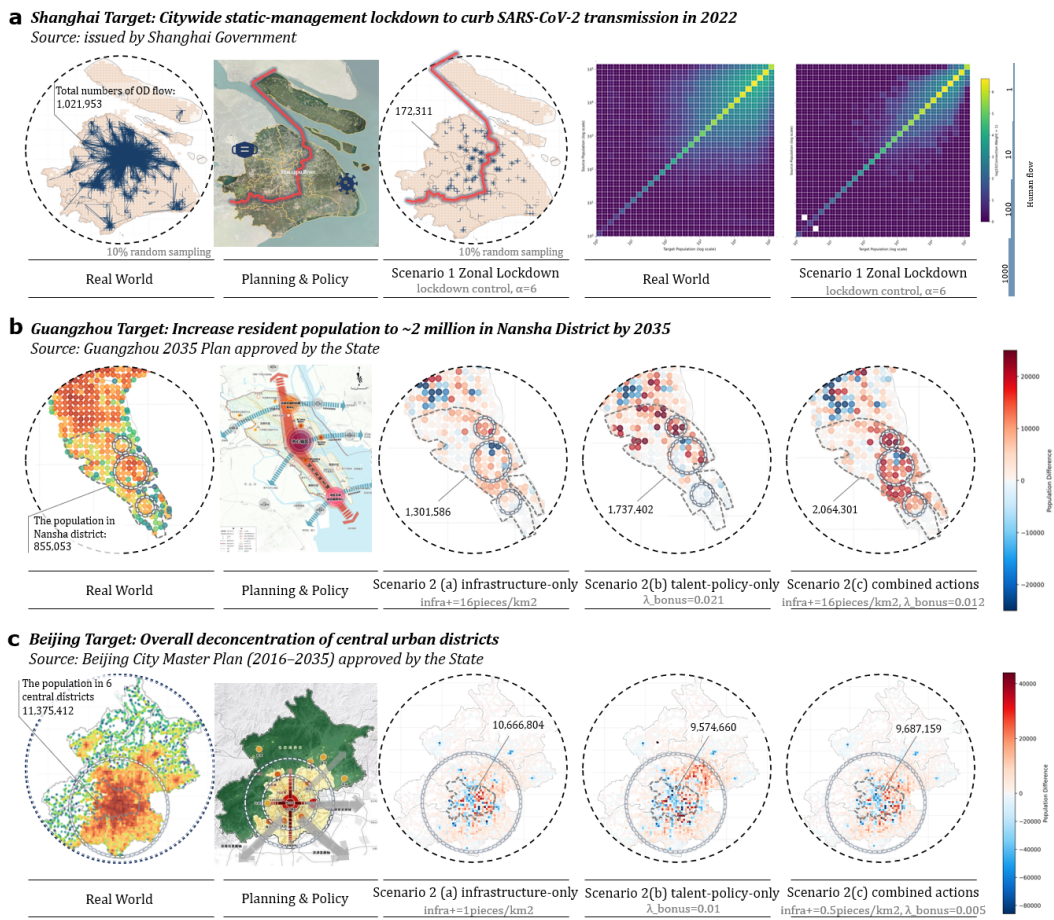

**Figure 5. Planning and policy scenarios. a**, Covid-19 lockdown in Shanghai(China). the first column is OD flows in daily routine, while the third is that after lockdown policy. The second column is the policy that prohibits the mobility across Huangpu River. The forth and fifth column is the population-mobility heatmaps before and after the policy. **b-c**, Nansha new area (Guangzhou, China) establishment and population dispersal in Beijing center (China). The first column is the population spatial distribution in 2023-2024. The second column is plan from government. The rest of the columns are population change under different planning measures. "infra" refers to adding infrastructure in specific area. "$\lambda\_bonus$" refers to the strength of talent policy. Blue dots mark ecological redline zones where high-density development is prohibited.